\documentclass[preprint]{jpsj2}
\title{Can Neural Networks Recognize Parts?}

\author{\textsc{Koji Matsumura}$^{1}$\thanks{E-mail address:
matsu@phys.chuo-u.ac.jp} and  
\textsc{Y-h. Taguchi}$^{1,2}$\thanks{E-mail address: tag@granular.com} }

\inst{$^{1}$Department of Physics, Faculty of Science and Technology, Chuo
University, 1-13-27 Kasuga, Bunkyo-ku, Tokyo 112-8551, Japan  \\
$^{2}$ Institute for Science and Technology, Chuo University, 1-13-27 Kasuga,
Bunkyo-ku, Tokyo 112-8551, Japan }

\abst{We have demonstrated neural networks can recognize parts by visual images.
Input signals are gray scale photographs of objects 
consisting of some parts and
output signals are their shapes.
By training neural networks by a few set of images,
without any supervision they become to be able to 
recognize the boundary between parts. 
}

\kword{Neural network, visual intelligence, recognition of parts}

\begin{document}
\maketitle

\section{Introduction} 

Visual intelligence (VI)\cite{VI} 
plays very important roles in the visual cognition.
In retina  system, we can accept only two dimensional projections of three
dimensional objects. Without any other informations, we have to recognize
three dimensional object from it.
Of course, there are infinitely many interpretations of
this two dimensional image received, but usually we reconstruct
unique three dimensional world. And it is often the proper
interpretation (otherwise, we would be extinct).

VI provides us the set of rules of interpretation to have
these proper reconstructions of three dimensional space.
There are many tasks to be solved by VI, for example,
reconstruction of roughness from gray scale image\cite{Knill_1991},
recognition of 	depth from line drawings\cite{Necker_1832}, and
decision of motion from sequential still images\cite{Exner_1875}.

One of such tasks is to recognize parts\cite{Bennett_1987}.
When we view a pair of iron dumbbells, we recognize it as two spheres
connected by a rod. Although there are some theories
\cite{Bennett_1987} to explain how
we can divide a pair of iron dumbbells into three parts,
there are no theories about how we can learn rules suggested by
these theories.
In this paper, we demonstrate that even a set of simple
neural networks can become to be able to recognize parts
without any supervision if many enough number of combinations 
of parts are presented, 
even if there are no informations about what each part is.
It seems to be very easier process than imagined.

In  \S \ref{sec2}, we have defined the objects from which we generate
visual images. Section \ref{sec3} 
describes how to train neural networks
so that it recognizes three dimensional shapes and parts
from the visual images. Discussions and Conclusions are
in \S \ref{sec4} and \S \ref{sec5}, respectively.

\section{Objects used}
\label{sec2}

\begin{figure}
(a)\includegraphics[scale=0.1]{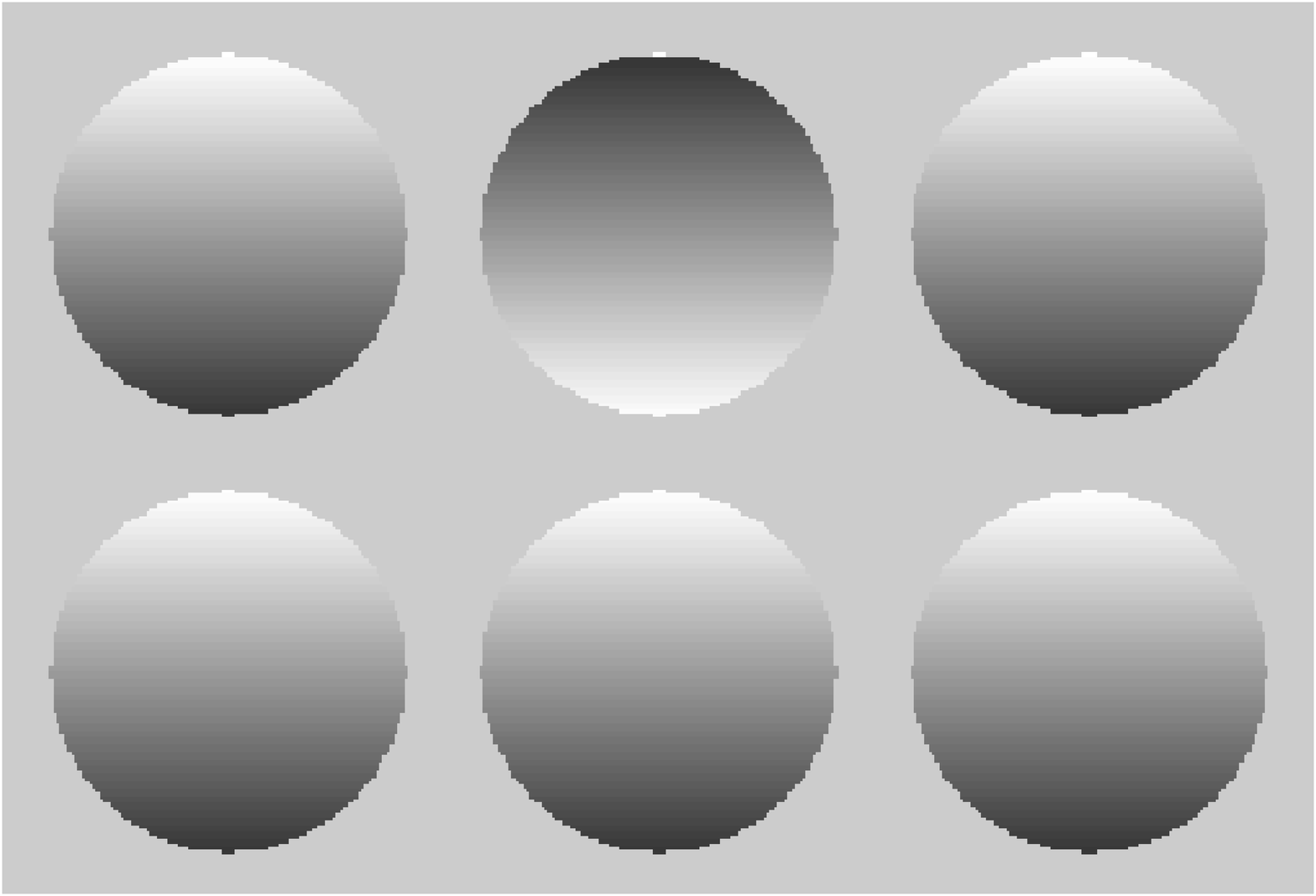}
(b)\includegraphics[scale=0.1]{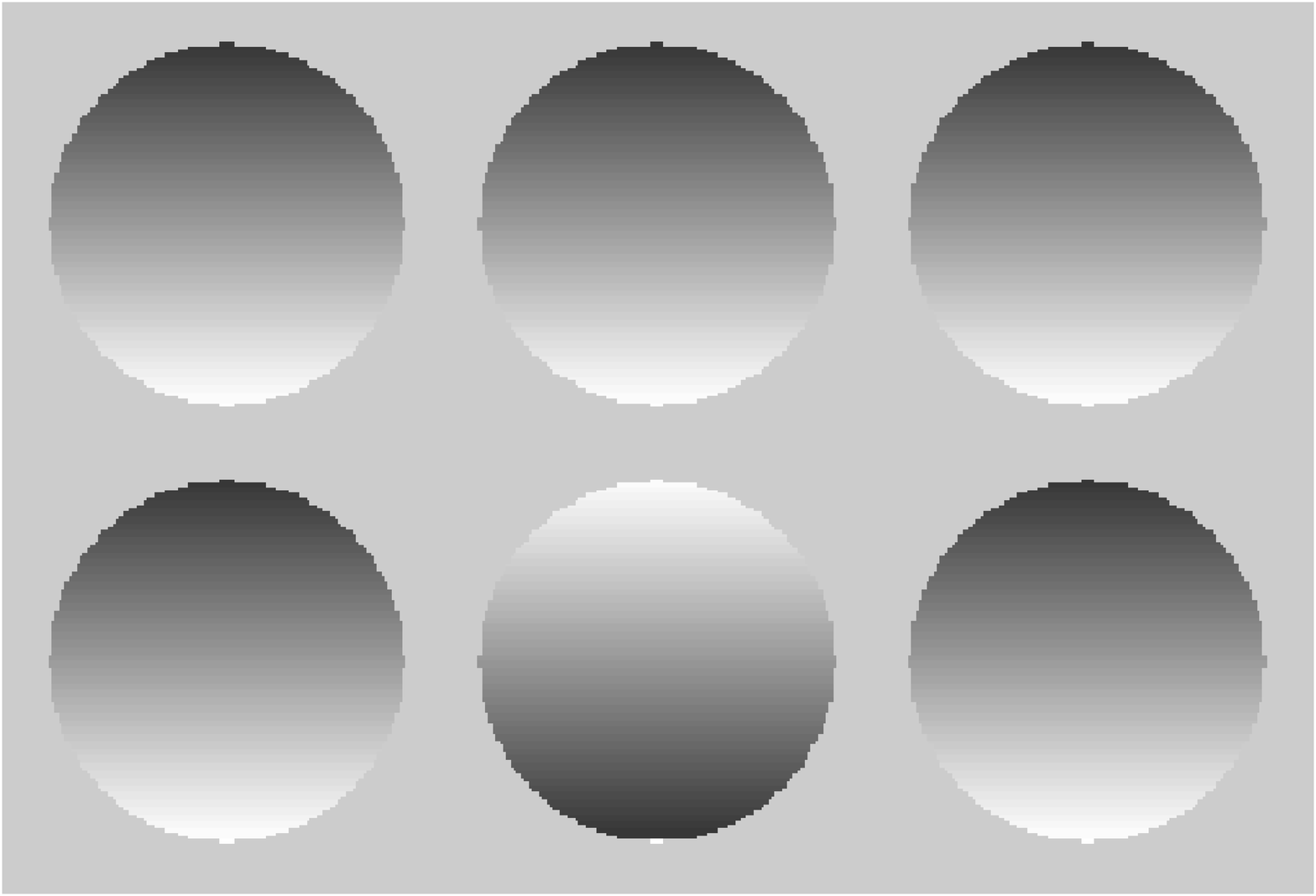}
\caption{Examples of images used for training}
\label{fig:objects}
\end{figure}

In order to make neural networks learn what the parts are,
we have to present grey scaled images of three dimensional objects.
However, if the objects are too complicated,
training  neural networks to learn them is simply time consuming.
It is a waste of time.
We need some simple images which are two dimensional projection of a
set of three dimensional objects and are easily recognized as a 
set of parts by human beings.
As such examples, we employ the images shown in Figs. \ref{fig:objects}.
If someone asks ``What does Fig. \ref{fig:objects}(a) look like?'',
the answer may be 
``Five hemispheres on a plate with a round hollow''.
These ``hemispheres'' and ``a hollow'' are the parts.
It is very easy for us to recognize these parts.
But how did we become to be able to do this?

\section{Training neural networks}
\label{sec3}

In order to check how easy it is to learn what the parts are,
we try to train neural networks to recognize them.
The neural networks used are standard three layered perceptrons,
\begin{eqnarray}
\mu_j &=& \sum_{i=1}^L a_{ij}x_i + a_{0j},\\
y_j &=& f(\mu_j), \\
\nu_k &=& \sum_{j=1}^M b_{jk}y_j + b_{0k},\\
z_k &=& f(\nu_k),
\end{eqnarray}
where
\begin{equation}
f(\mu)=\frac{1}{1- \exp(-\mu)},
\end{equation}
and $x_i,y_j$ and $z_k$ are  values of the input neurons, the neurons in
the hidden layer and the output neurons respectively.
$a_{ij}$s and $b_{jk}$s are connection coefficients
which are trained by usual back propagation procedure.

\begin{figure}
\includegraphics[scale=0.5]{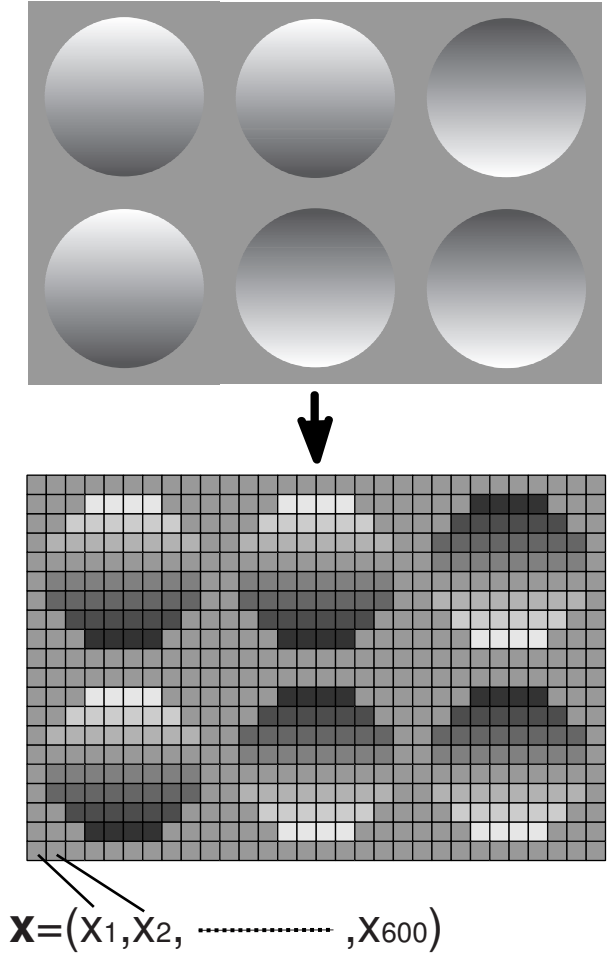}
\caption{Inputs $x_i, (i=1,..,600)$}
\label{fig:lattice}
\end{figure}

In order to decide values of  input $x_i$s, we have subdivided a image
into $20 \times 30$ lattices
(Fig. \ref{fig:lattice}). $x_i (i=1,..,600)$ takes 
1(0) if center pixel is white (black).
In total, $2^6 = 64$ images can be considered 
because each of the six parts has two possibilities that it can take.

\subsection{Recognition of a hollow or a hemisphere}
\label{sec:hollow_shape}
First, we would like to check whether neural networks 
can recognize both a hollow and a hemisphere successfully.
Thus, we define six $z_k, (k=1,..,6)$ as follows,
while suffix $k$ corresponds to one of six parts;
if the $k$th part is a hemisphere (a round hollow), $z_k$ takes 1(0).
We have employed 600 neurons in the hidden layer.
Although one may think that it is too large for this simple task,
it is not the case
because later we use this for learning the three dimensional shapes.

\begin{figure}
\includegraphics[scale=0.3]{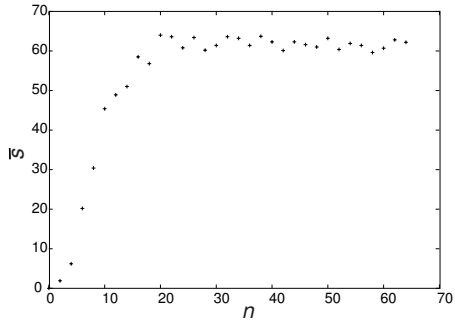}
\caption{Averaged number of correctly recognized patterns
out of total 64 images as a function of number of
trained images. (a hollow or a sphere recognition)}
\label{fig:training}
\end{figure}

In Fig. \ref{fig:training}, 
we have shown 
the dependence of 
the average number of
patterns $\bar{S}$ recognized correctly
by trained neural networks 
upon a number $n$ of images 
used for training.
Of course, $z_k$s take non integer values between 0 and 1,
but we regard $z_k=1(0)$ when $z_k >(<)0.5$.
Averages are taken over ten independent training  for
each $n$. As can be seen easily, if $n$ is larger than one third of
total number of images, neural networks correctly recognize  hollows and
 hemispheres for all images. Thus, neural networks can
recognize a hollow and a hemisphere correctly.

\subsection{Recognition of 3D shapes}

\begin{figure}
\includegraphics[scale=0.5]{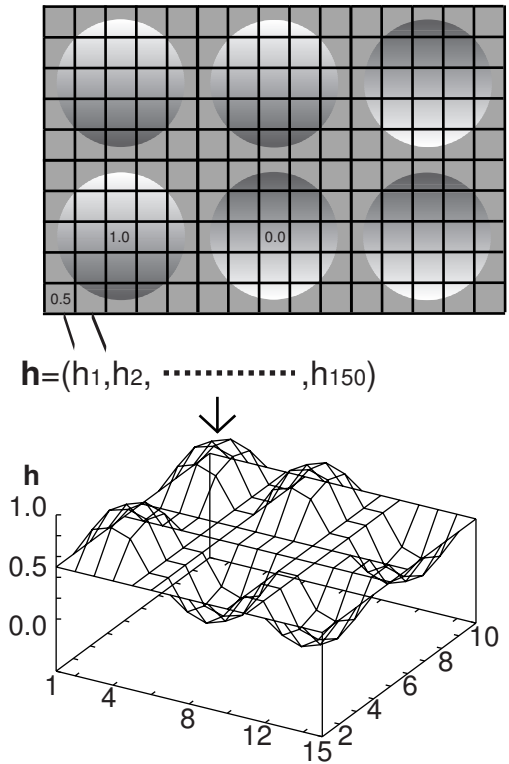}
\caption{Output $h_i$s for three dimensional shape recognition.}
\label{fig:3d}
\end{figure}

Next we try to make neural networks recognize three dimensional shapes.
This time, outputs $z_k$s are the coarse grained height $h_k$ of a hollow or a
hemisphere (Fig. \ref{fig:3d}).
We have subdivided surface of three dimensional shapes into
$15 \times 10 =150$ lattices. $h_k$ takes value between 1 and 0.
The surface of flat plate is regarded to have height 0.5,
and the bottom of hollows has 0 and the top of hemisphere has 1.0.

\begin{figure}
\includegraphics{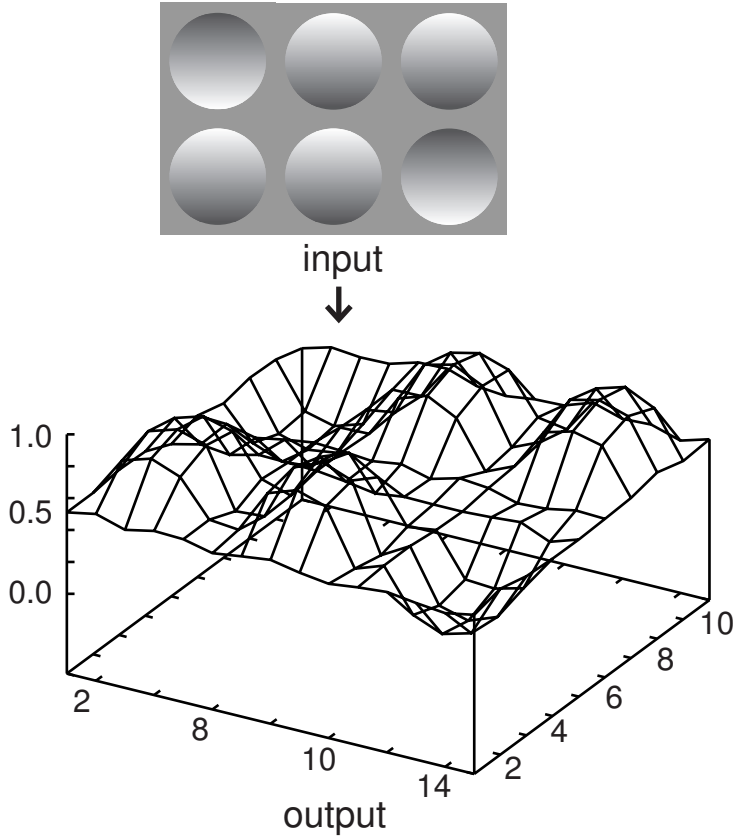}
\caption{Output $h_i$s by trained neural networks.
Input is an unknown (not used for training) image.}
\label{fig:trained_3d}
\end{figure}

In Fig. \ref{fig:trained_3d}, we have shown the
ability of trained neural networks.
This neural networks are trained using 24 out of total 64 images.
Then a image not used for training is presented.
As can be seen easily, the neural networks can easily recognize
the 3D shape even if unknown image is presented.

\begin{figure}
\includegraphics[scale=0.3]{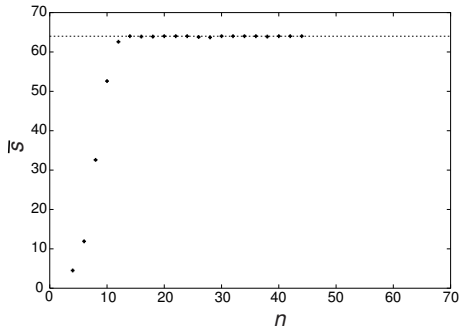}
\caption{Averaged number of correctly recognized patterns
out of total 64 images as a function of number of
trained images. (3D shapes recognition)}
\label{fig:3d_shape_training}
\end{figure}

In Fig. \ref{fig:3d_shape_training}, 
we have shown 
the dependence of 
the average number of
patterns $\bar{S}$ recognized correctly
by trained neural networks 
upon a number $n$ of images 
used for training.
It
is possible for neural networks to
learn 3D shapes if $n$ is larger than 15.
Thus, neural networks correctly recognize 3D shapes.

\subsection{Recognition of parts}

\begin{figure}
\includegraphics[scale=0.5]{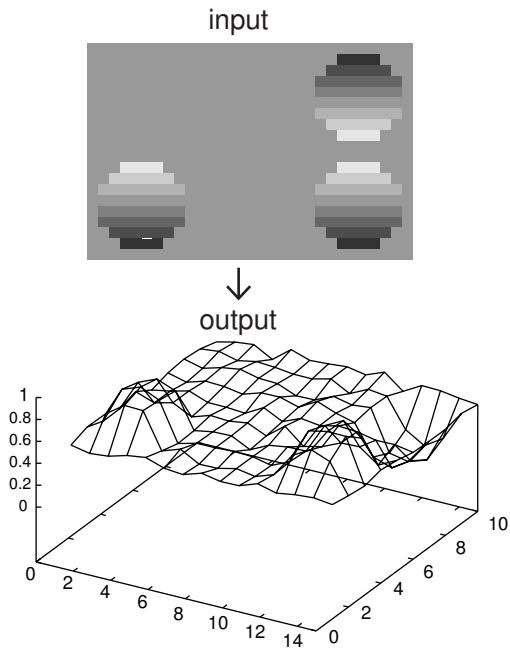}
\caption{Recognition of parts}
\label{fig:parts}
\end{figure}

Until now, we did not provide any information about
what the parts are. However, neural networks have learned it as shown below.
In order to see whether the neural networks 
recognize parts, we have shown three hollows/spheres on a flat
plane to the neural networks. 
If they can recognize what the parts are, they can
reproduce 3D shapes.
As shown in Fig. \ref{fig:parts},
neural networks can recognize each hollow/hemisphere as a part.
Even if there is only one hollow/hemisphere on a plate,
they can reproduce three dimensional shapes correctly.
This means that
without any supervisions, 
neural networks can recognize each hollow/hemisphere as a part.

\section{Discussion}
\label{sec4}

How do neural networks relate the regions of a grey scale image to
the regions on a flat plate? 
We did not provide such a information at all.
However, once neural networks recognize correspondence
between parts in images (input information) and parts
in 3D shapes (output information), it is essentially
to find relations between 6 bit input and 6 bits output
(In bit interpretation for example, a hemisphere corresponds to 1 and
a hollow corresponds to 0.). Thus, it is a very easy task for neural networks.

\begin{figure}
(a)\includegraphics[scale=0.3]{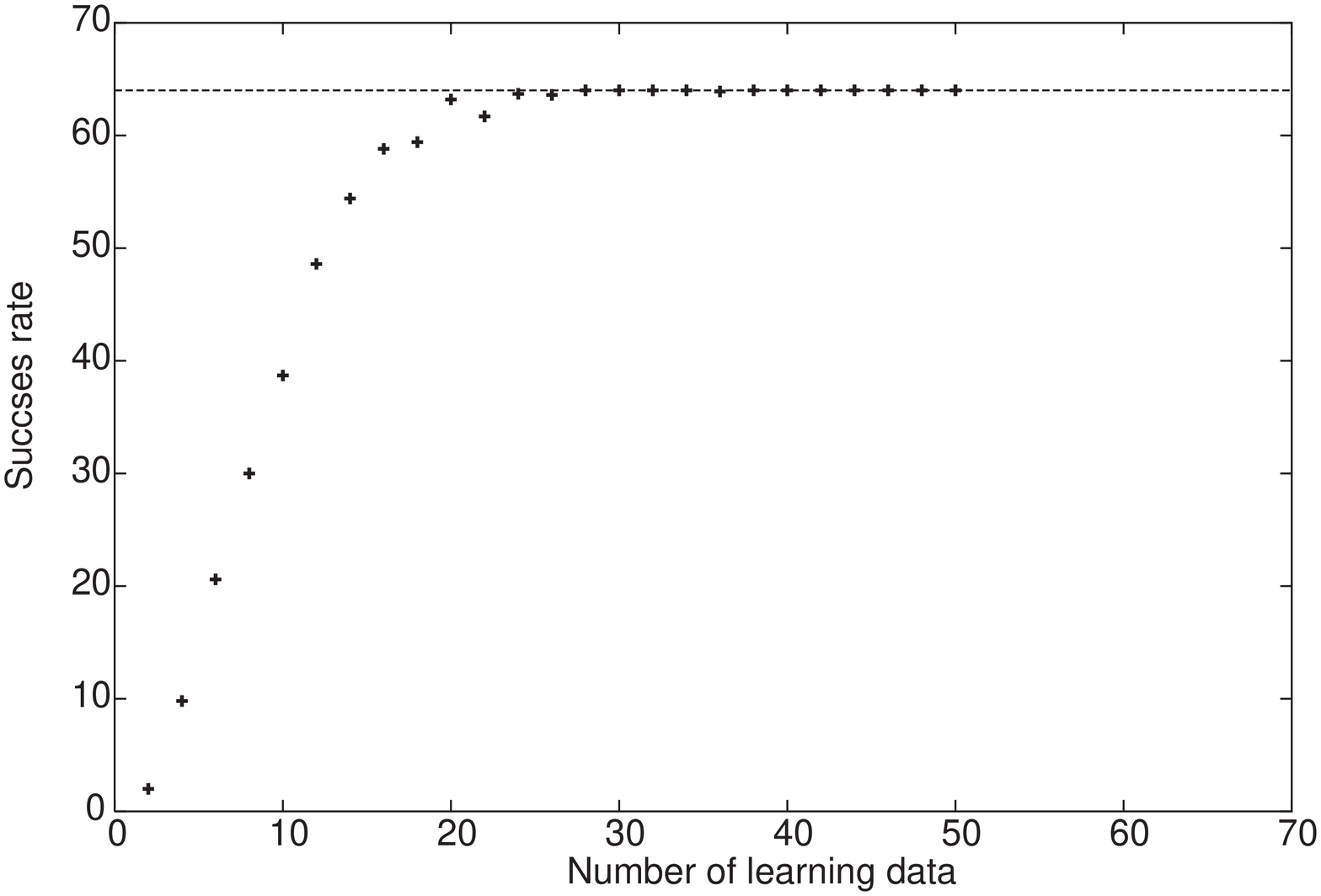}
(b)\includegraphics[scale=0.3]{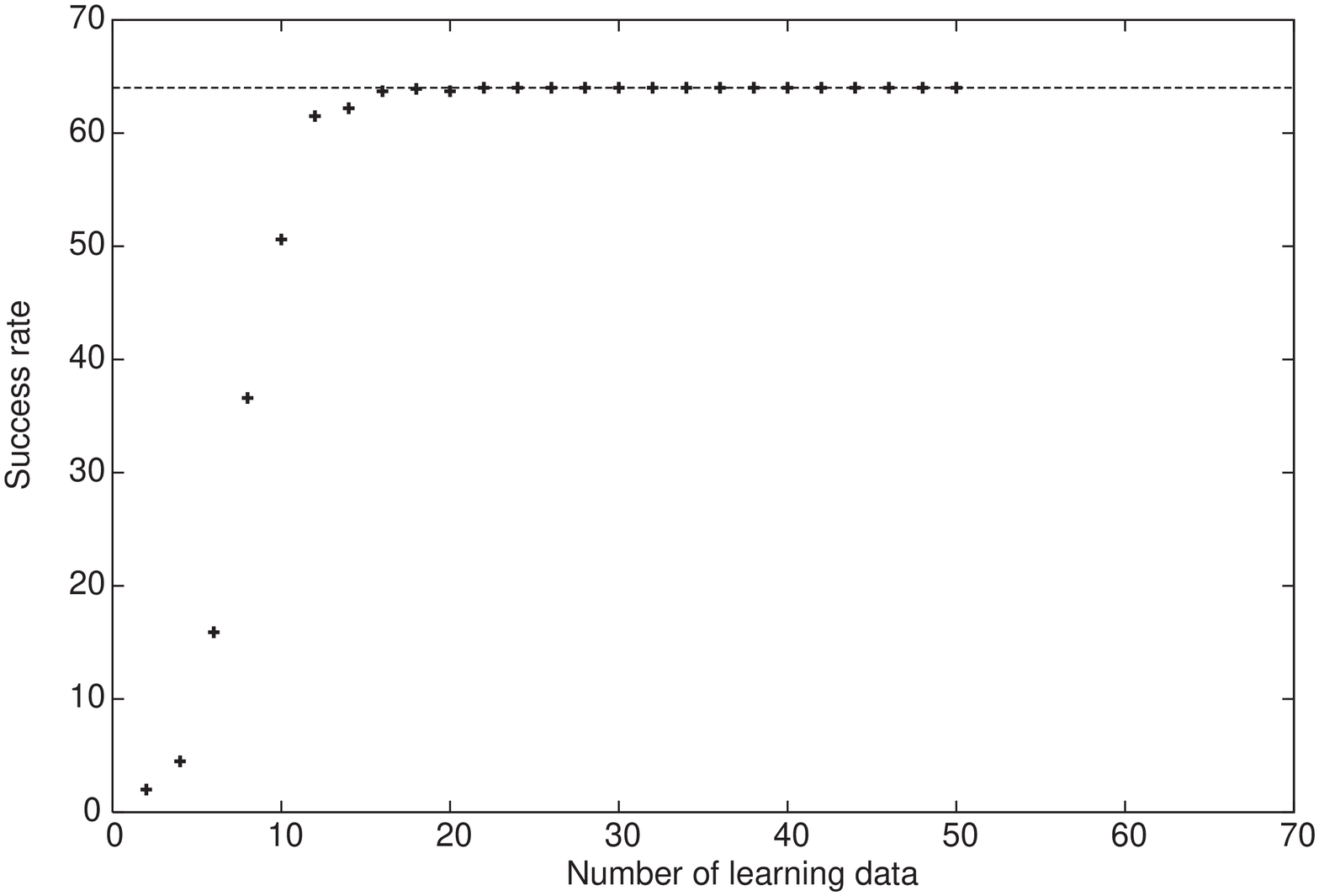}
\caption{(a) Averaged number of correctly recognized patterns
out of total 64 images as a function of number of
trained images (For 6 input/output neurons).
(b) The same as (a) for 3D shape recognition with
50 neurons in hidden layers}
\label{fig:onetoone}
\end{figure}

In order to check the easiness,
we use 6 neurons  at input and output layers and 30 neurons at hidden layers.
$x_i$ and $z_k$ take 0 or 1 and neural networks are trained such that $x_i=z_k$
when $i=k$.
As shown in  Fig. \ref{fig:onetoone}(a),
it is possible for neural networks to do this.
Thus, at maximum, neural networks need only 30 neurons in hidden layer.
Thus, if neural networks recognize parts, the numbers of neurons
can be  as small as 600.

Actually, as shown in Fig. \ref{fig:onetoone}(b),
neural networks can have the same ability as Fig. \ref
{fig:3d_shape_training} even if
the number of neurons is only 50. 
This is almost the number of neurons in the hidden layer
of neural networks whose ability is shown in Fig. \ref{fig:onetoone}(a).
Thus we can conclude that neural networks recognize parts well and
can drastically reduce the number of neurons in hidden layers.
This is how to learn what the parts are.
If the simple network can recognize it so easily,
our neuron can do the same without difficulty.
This may be the reason why we became to be able to recognize the image
as a set of parts.
It can reduce the number of neurons in hidden layers
drastically as expected. Without recognition of parts, it is impossible to
reduce the number of neurons in the hidden layer.

\section{Conclusion}
\label{sec5}

In conclusion, we have shown that neural networks
can divide the image into parts automatically
during training process. 
It turns out to reduce number of used neurons in hidden layer,
i.e., memories drastically. This may be the reason why
we became to be able to recognize the image as a set of parts.



\begin{thebibliography}{99}
\bibitem{VI} D. D. Hoffman: {\it Visual Intelligence},
(W. W. Norton \& Company, New York, 1998).
\bibitem{Knill_1991} D. C. Knill and D. Kersten:
  Nature, {\bf 351}, (2001), 228.
\bibitem{Necker_1832} L. A. Necker:
 Phil. Mag., {\bf 3} (1832) 329.
\bibitem{Exner_1875} S. Exner:
S\"{u}zungsberichte Akademie Wissenschaft Wien,
{\bf 71} (1875) 156.
\bibitem{Bennett_1987} B. M. Bennett and D. D. Hoffman:
in W. A. Richards and S. Ullman (Eds.) {\it Image
understanding 1985-1986}, (Ablex, New York, 1987), 215. 
\end{thebibliography}
\end{document}